\magnification=\magstep1
\baselineskip = 16pt

\font\title = cmr10 scaled 1440

\def\<{\langle}
\def\>{\rangle}

\def\Pphi{|\phi\>\<\phi|}

\def\fr#1/#2{{\textstyle{#1\over#2}}} 

\newcount\ftnumber
\def\ft#1{\global\advance\ftnumber by 1
          {\baselineskip 13pt    
           \footnote{$^{\the\ftnumber}$}{#1}}}

\newcount\eqnumber

\def\equ(#1){
    \ifx\DRAFT\undefined\def\DRAFT{0}\fi	
    \global\advance\eqnumber by 1%
    \expandafter\xdef\csname !#1\endcsname{\the\eqnumber}%
    \ifnum\number\DRAFT>0%
	\setbox0=\hbox{#1}%
	\wd0=0pt%
	\eqno({\offinterlineskip
	  \vtop{\hbox{\the\eqnumber}\vskip1.5pt\box0}})%
    \else%
	\eqno(\the\eqnumber)%
    \fi%
}
\def\(#1){(\csname !#1\endcsname)}
           
\def\vN{von Neumann}
\def\VN{Von Neumann}
\def\cR{{\cal R}}
\def\cS{{\cal S}}
\def\cT{{\cal T}}

\def\cW{{\cal W}}

\def\Exp{{\rm Exp}}
\def\Tr{{\rm Tr}}
\def\qm{quantum mechanics}
\def\se{subensemble}

\def\ni{\noindent}
\def\llra{\longleftrightarrow}
\def\'{$'$}

{\title Homer nodded: von Neumann's surprising oversight}
\bigskip
\centerline{N. David Mermin}

\centerline{Laboratory of Atomic and Solid State Physics}

\centerline{Cornell University, Ithaca NY 14853}
\bigskip
\centerline{R\"udiger Schack}

\centerline{Department of Mathematics, Royal Holloway University of London}

\centerline{Egham, Surrey TW20 0EX, UK}

\bigskip

\bigskip
\noindent{\sl Abstract.}  
We review  the famous no-hidden-variables theorem  in John \vN's 1932 book on the mathematical foundations of \qm.   We describe the notorious gap in \vN's argument, pointed out by Grete Hermann in 1935 and, more famously, by John Bell in 1966.  We 
disagree with recent papers claiming that Hermann and Bell failed to understand what \vN\ was actually doing.

\bigskip

\noindent {\bf 1.  Introduction}
\medskip

Over half a century ago John Bell, 1966,  criticized 
the famous argument of John \vN, 1932,  that hidden-variable theories cannot underlie quantum mechanics.    Unknown to Bell, Grete Hermann, 1935,  had published the same criticism three decades earlier.  Bell then went on to prove an important  no-hidden-variables theorem of his own,\ft{Not to be confused with the more famous ``Bell's theorem'' of Bell, 1964.  The relation between these two different theorems of Bell is discussed in Mermin, 1993.}  without 
making the mistake of \vN\ that he (and Hermann) 
had noted.\ft{Bell  raises a similar objection to his own  
improved argument, but this is not relevant to our concerns here.}

Recently Jeffrey Bub, 2010,  claimed that Bell had misunderstood \vN's argument, and quite recently Dennis Dieks, 2017,  expanded on Bub, adding similar criticism of the earlier work of Hermann.  
We, however, agree with Hermann's and Bell's reading of  \vN, and believe that   
Bub and Dieks  fail to make sense of  the surprising gap in
 \vN's argument that Hermann and Bell correctly identified.\ft{Bub, 2011, adds Mermin, 1993, to his list of those who read \vN\ wrong.}  

In Section 2 we summarize  \vN's argument against hidden variables, and identify his oversight.       In Section 3 we describe Bell's criticism of \vN's argument. While  Bell does not convey some of \vN's subtle distinctions,  he does get \vN's error exactly right.    Section 4 describes the much earlier, but less well-known criticism of \vN\ by Hermann.  She captures better than Bell  the full character of \vN's argument, and, like Bell, correctly explains what's wrong with it. 

We comment in Sections 2-4 on Bub's and Dieks' reading of \vN\ and why we believe that reading is wrong.

\bigskip
\noindent {\bf 2. \VN's argument}
\medskip
\centerline{\sl a.  \VN's assumptions}
\medskip
\VN\ derives much of the structure of quantum mechanics together with his argument against hidden variables, from four assumptions.    Because the four assumptions lead not only to the structure of quantum mechanics, but also to \vN's no-hidden-variables argument,  if hidden variables are nevertheless compatible with quantum mechanics, then at least one of his assumptions must be wrong.  \VN\  concludes that one cannot construct a hidden-variables model without doing irreparable damage to the structure of quantum mechanics.  But Hermann and Bell both point out that one of \vN's four assumptions, essential for the no-hidden-variables part of his argument, can be dropped without altering the structure of ordinary quantum mechanics (implied by the remaining three) in any significant way.  

Two of \vN's assumptions,  $A'$ and $B'$, deal with ``physical quantities'' and their measurement.   They are about statistical properties of data,  and they make no explicit reference  to the formalism of \qm.   
The other two assumptions,  I and II, make no explicit mention of measurement, data, or statistics.   They simply associate physical quantities  with Hermitian operators on a Hilbert space,  in a way that preserves certain structural relationships obeyed by both the physical quantities and the Hermitian operators, thereby bringing into the story much of the formal mathematical apparatus  of \qm.    Here are \vN's four assumptions:\ft{We give them the names used by \vN.}  
\medskip

\noindent {\bf Assumption A}$'$:  (p.~311\ft{Page references are all to the English translation of \vN.})   There exists an expectation function \Exp\ from physical quantities to the real numbers.

A physical quantity $\cR$ can be subject to a measurement, which yields a real number $r$.    If you have an ensemble of physical systems, all associated with the same set of physical quantities, and you measure the same physical quantity $\cR$ on a  large enough random sample of the systems, then the mean of all those measurement outcomes is called $\Exp(\cR)$.\ft{In 1932 most physicists talked about probabilities in terms of ensembles. In this paper we follow von Neumann's language. For a given physical system an ensemble can be understood as an assignment of probabilities to the outcomes of all possible ways to measure physical quantities (or sets of jointly measurable physical quantities) defined on that system.  For a given ensemble, the function Exp($\cR$) is then the standard expectation value for the probability distribution associated with the particular measurement of $\cR$.}  
Implicit in Assumption A$'$, and in the notation Exp($\cR$), is the physical assumption, not always emphasized,  that 
this mean value does not depend on which of several possible distinct ways of measuring $\cR$ might be chosen.  

{\it One\/} way to define a  physical quantity is to specify a way to measure it.    As an important example,   if $\cR$ is a physical quantity that one does know how to measure, and  $f$ is a function that takes real numbers to real numbers, then one can define another physical quantity $f(\cR)$ by specifying that  to measure $f(\cR)$  you measure $\cR$ and then apply $f$ to the outcome $r$ of the $\cR$-measurement.

We shall point out below that the criticisms of Hermann's and Bell's readings of \vN\ by Bub and Dieks are invalidated by the fact that \vN's four assumptions also provide {\it another\/} way to  define physical quantities that makes no explicit mention of measurements.   

Assumption A\' also states explicitly that Exp($\cR$) is  non-negative if the physical quantity $\cR$ is ``by nature'' non-negative.  Nobody has any issues with this.  

\bigskip

\noindent {\bf Assumption B}$'$:  (p.~311) If $\cR, \cS, \ldots$ are arbitrary physical quantities, not necessarily simultaneously measurable, and $a, b, \ldots$ are real numbers then the expectation function Exp is linear:

 $$\Exp(a\cR + b\cS+ \cdots) = a\,\Exp(\cR) + b\,\Exp(\cS)+ \cdots.\equ(B')$$
\bigskip

 If several different physical quantities $\cR, \cS,\ldots$  can be simultaneously measured, then you can define a physical quantity that is a function $f$ of them all by specifying that $f(\cR, \cS\ldots)$  is measured by measuring them jointly, and   
applying $f$ to the results $r, s,\ldots$ of all those measurements.   The linearity condition B\'   for jointly measurable quantities follows straightforwardly from this definition, applied to the function $f(r, s, \ldots) = ar + bs +\cdots.$ 

Now it is one of the most important features of quantum mechanics that not all physical quantities {\it can\/} be simultaneously measured.\ft{The acknowledgment that some physical quantities cannot be jointly measured introduces a crucial feature of quantum mechanics even into assumptions A\' and B\'.}   
Extending the scope of Assumption B\' to quantities $\cR, \cS,\ldots$ that are {\it not\/} jointly measurable is problematic, however, since at this stage it is not even clear what  $a\cR + b\cS+\cdots$ in B\' might mean for such quantities.  Indeed,
  \vN\  immediately remarks that B\' characterizes such a linear combination ``only in an implicit way'', since there is ``no way to construct from the measurement [instructions] for $\cR, \cS,\ldots$ such [instructions]  for $\cR + \cS +\cdots.$"\ft{Bottom of p.~309.}  
  
  Bub and Dieks both take this to mean that \vN\ uses assumption B\'  to {\it define\/} linear combinations of physical quantities that are not simultaneously measurable.  This is the entire basis for their criticisms of Bell and Hermann. If B\' is just a definition, it cannot also be an invalid assumption, as Hermann and Bell maintain.  But as we shall see below,  the full set of \vN's four assumptions contains another  way to define linear combinations of physical quantities that are not simultaneously measurable.  With that alternative definition, Assumption B\' can indeed impose a nontrivial constraint on the values an Exp function can have for such linear combinations.  There is no reason to insist that  Assumption B\' must be taken as a definition. 
 
 \bigskip
\noindent {\bf Assumption I}:    (p.~313)  There is a 1-to-1 correspondence between physical quantities $\cR$ and Hermitian operators R that act on a Hilbert space.  For any real-valued function $f$, if the quantity $\cR$ has the operator $R$, then the quantity $f(\cR)$ has the operator $f(R)$.

$$  \cR \longleftrightarrow R\ \Longrightarrow\ f(\cR) \longleftrightarrow f(R). \equ(I)$$

The requirement  that this 1-to-1 correspondence must be preserved by functions is quite powerful.   
We have noted in our discussion of Assumption A\'  \vN's specification of how to define functions of a  physical quantity.   Standard Hilbert space mathematics tells us how to define functions of a Hermitian operator.   Requiring,  as  Assumption I does, that these two quite different ways of evaluating functions should preserve the one-to-one correspondence between physical quantities and Hermitian operators has surprisingly strong consequences.   Appendix 1   illustrates the power of this function-preserving 1-1  correspondence.
 
Because this association of physical quantities with Hermitian operators is  {\it one-to-one\/}, it is possible to use Hermitian operators to {\it define\/} physical quantities, and vice-versa.   Assumption II provides a  pertinent example of this.

\bigskip  

\noindent {\bf Assumption II}:  (p.~314) If the physical quantities $\cR, \cS,\ldots$ have the Hermitian operators $R, S,\ldots$, then the physical quantity $a\cR + b\cS +\cdots$ has the Hermitian 
operator $aR + bS + \cdots$, whether or not $\cR, \cS, \ldots$ are simultaneously measurable:\ft{\VN's statement of Assumption II is only for the special case $a=b=1$.   But Assumptions I and A\' tell us that $aR$ is the Hermitian operator associated with the physical quantity $a\cR$, which leads directly to the more general form we give here.} 

$$a\cR + b\cS +\cdots\ \longleftrightarrow\ aR + bS +\cdots.\equ(II)$$

Assumption II provides the obvious way to  define  $a\cR + b\cS +\cdots$ for 
sums of physical quantities that are not simultaneously measurable.   There is no problem in defining linear combinations  of arbitrary Hermitian operators.    
The physical quantity $a\cR + b\cS\ +\cdots$ can then be {\it defined\/}, under Assumption II, to be the one that corresponds to the Hermitian operator $aR + bS +\cdots$, where $R, S,\ldots$ are the Hermitian operators that correspond to the individual physical quantities $\cR, \cS,\ldots$.  This definition reduces  to the simple definition in terms of measurement outcomes when the quantities are jointly measurable.
Assumption II extends that definition when they are not.    

This observation invalidates what Bub and Dieks have to say about Hermann's and Bell's alleged misunderstanding of \vN.  Whether \vN\  {\it intended\/} to define such sums through Assumption II is beside the point, though we believe he did,\ft{\VN's derivation of the density matrix form for the Exp-function, mentioned  below, makes explicit use of Assumption II to construct such linear combinations, to which he then applies Assumption  B\'.}  and Hermann clearly thought that he did.   To invalidate Bub's and Dieks' criticism of Hermann and Bell it is enough that an alternative definition {\it exists\/} in addition to the definition Bub and Dieks attribute to \vN.\ft{In what follows we expand on how \vN\ uses his four assumptions to arrive at his no-hidden-variables theorem, and why his conclusions that hidden variables would undermine the fundamental principles of quantum mechanics are indeed not justified by his argument.}    

\vfil\eject
\medskip
\centerline{\sl b.  What \vN\ proves with his assumptions.}
\medskip

\VN\ first proves\ft{Pages 314-316, with some additional mopping up on pps.~316-320.} that if an ensemble of physical systems   and the associated Exp function satisfy all four of his assumptions, then the Exp function  for that ensemble must have the form $$\Exp(\cR) = \Tr(UR),\equ(Tr)$$ where $U$ is a non-negative\ft{\VN\ uses the term ``definite''.} Hermitian operator characteristic of  the ensemble but  independent of the physical quantity $\cR$.   In modern language there must be a density matrix $U$, such that the Exp function for the ensemble is the trace of the product of that density matrix with the Hermitian operator that corresponds to that physical quantity.\ft{\VN's proof is quite straightforward.    Three decades later Andrew Gleason proved what is now known as Gleason's Theorem: that the density matrix form \(Tr) follows  from premises essentially equivalent to Assumptions A\', I, and II.  Gleason does not use assumption B\' for physical quantities that are not jointly
measurable. His argument is notoriously intricate (and requires that the Hilbert space have three or more dimensions).} 

The Exp function characterizing a pure quantum state $\phi$ is indeed of the form \(Tr) with the density matrix $U$ given by $\Pphi$.   And, of course, the ensembles associated with ordinary quantum states do indeed satisfy all four of \vN's assumptions.

\VN\  addresses the question of hidden variables on p.~323.\ft{He also brings up the question  on pps.~209-210, but defers answering it, promising to show later that ``an introduction of hidden parameters is certainly not possible without a basic change in the present theory.''}  He asks whether the dispersion of any 
ensemble characterized by a wave function $\phi$ could result from the fact that such pure states are not the fundamental states, but only statistical mixtures of several more basic states.         To specify such ``actual states'' one would need additional data --- ``hidden parameters'', which we denote here collectively by $\lambda$.    When adjoined to the quantum state $\phi$ these hidden parameters would determine everything --- i.e. the resulting subensembles would be free of dispersion: $$\Exp_{\phi,\lambda}(\cR^2) = (\Exp_{\phi,\lambda}(\cR))^2\equ(nodis)$$ for all physical quantities $\cR$.  The statistics of the  nondeterministic ensemble, characterized by \(Tr) with  $U = U_\phi = \Pphi$,  would result from appropriately weighted averages over all the actual states, ($\phi,\lambda$),  into which the $\phi$-ensemble was decomposed by the hidden parameters.\ft{Putting it in terms of probability distributions $p$ rather than ensembles, the question is whether $p_{[\phi]}$ can be expressed as a weighted average of dispersion-free distributions,  conditioned not just on the state $\phi$, but also on the additional parameters $\lambda$.}  

\VN\ shows (again straightforwardly) that a $\phi$ ensemble cannot be so decomposed into dispersion-free ($\phi,\lambda$) \se s provided the Exp functions for the \se s, $\Exp_{\phi,\lambda}$, are also of the form \(Tr) with density matrix $U$ given by some $U_{\phi,\lambda}$.  Therefore if the Exp functions for quantum states {\it can\/} be represented by weighted averages of Exp functions for dispersion-free subensembles, then some of those subensembles cannot have Exp functions of the form \(Tr), and therefore some of \vN's four assumptions must fail for some of those subensembles.   

Which assumptions might it be that fail for the dispersion-free \se s?

 \medskip
\centerline{{\sl c.  \VN\ nods.}\ft{``Homer nods.'':  {\sl Even the best of us sometimes slip up.\/}    From John Dryden's translation of  line 359 of Horace's {\it Ars Poetica: indignor quandoque bonus dormitat Homerus\/}.}}

\vskip 5pt

\VN\ clearly believes I and II to be the assumptions that must be abandoned if there are dispersion-free \se s.   
When he states  that ``the established results of quantum mechanics can never be derived'' (p.~324) if there are dispersion free \se s,'' the reason he offers is that if they did exist, then ``it [would be] impossible that the same physical quantities exist with the same function connections (i.e., that I and II hold).''   That is indeed what I and II are about --- functional relations among physical quantities, mediated by their corresponding Hermitian operators.  Assumptions I and II, as noted above,  make no mention of ensembles or statistical distributions.  They specify broad structural relations, that it might be reasonable to expect to hold for physical quantities, regardless of what subensembles they might be measured in.\ft{When \vN\ adds ``Nor would it help if there existed other, as yet undiscovered, physical quantities, in addition to those represented by the operators in quantum mechanics, because the relations assumed by quantum mechanics (i.e., I, II) would have to fail already for the by-now  known quantities," he underlines that he is blaming assumptions I and II.}

If indeed it was assumptions I and II that \vN\ expected to fail for the dispersion-free subensembles, then one can understand  his now notorious ``It is therefore not, as is often assumed, a question of a reinterpretation of \qm, --- the present system of \qm\ would have to be objectively false, in order that another description of the elementary processes than the statistical one be possible.''  (p.~325)

So strong a conclusion might indeed be appropriate if assumptions I and II were the only suspects.   But there are other suspects, A\' and B\' that \vN, unaccountably, fails to question.  These have to do with the nature of physical quantities and the statistics of ensembles.   They have nothing to do with  ``function connections''  among physical quantities,
 or ``relations assumed by quantum mechanics.''   Could assumptions A\' or B\' be sacrificed for the dispersion-free \se s without making ``the present system of quantum mechanics $\ldots$ objectively false"?
 
It might indeed be  radical to abandon for \se s the idea, A\',  that single physical  quantities  and simultaneously measurable sets give rise to statistics that do not depend on the particular way in which they are measured.   One could argue whether that would be more or less radical than abandoning I and II for the \se s.    But why bother to argue?    Why not simply give up assumption B\' for linear combinations of physical quantities that are not simultaneously measurable?  

It is a peculiar feature of ordinary quantum mechanics that Assumption B\'  
holds for the mean values over the $\phi$-ensembles specified by quantum states, even when the physical quantities cannot be jointly measured.    But there is no  compelling reason to expect that B\' should continue to hold for averages over the ($\phi,\lambda$)-\se s  into which the $\phi$-ensembles might be subdivided by specifying additional hidden variables.   

Bub and Dieks pass over B\',  as a candidate for the assumption that fails for the dispersion-free \se s, because they insist on interpreting it as nothing more than a definition.   Dieks says that it would make no sense to reject B\' for those 
\se s because it is ``analytic''.   But as  emphasized above,  Assumption II provides a powerful alternative way to define linear combinations of physical quantities that are not jointly measurable.    In terms of that definition it is not only meaningful  to reject B\' for the hypothetical dispersion-free \se s, but quite compatible with the general structure of ordinary quantum mechanics.
 Thanks to Hermann and Bell, Bub and Dieks are aware that they need  a reason for not blaming  B\'.   \VN, who was unable to benefit from Bell's later criticisms\ft{It would be interesting to know if he ever became aware of Hermann's.} seems just to have overlooked the possibility.    Homer not only nodded.  He seems to have been fast asleep.   Bell's describing his oversight as ``silly" in a magazine interview does not strike us as excessive.\ft{See Mermin, 1993.}    

There is no reason at all to require the Exp functions on possible dispersion-free \se s to be linear on physical quantities that are not simultaneously measurable.  Maintaining ``the established results of quantum mechanics'' only requires B\' to hold when those \se s are recombined to make up the $\phi$-ensemble characterizing the full quantum state $\phi$.  This is precisely the point made by John Bell fifty years ago, and, thirty years before Bell, by Grete Hermann.

\bigskip
\noindent {\bf 3. Bell's criticism of \vN.}
\medskip

The most important part of Bell, 1966, is his better version of \vN's attempt at a no-hidden-variables theorem.  Bell restricts \vN's assumption B\'  to physical quantities $\cR, \cS,\ldots$ that {\it can\/} be simultaneously measured.  The linear combination $\cW = a\cR + b\cS+ \cdots$ can then be measured by jointly measuring $\cR, \cS,\ldots$ and forming the corresponding linear combination of those measurement  outcomes.
With a more elaborate argument, quite different from \vN's, Bell can still rule out dispersion-free \se s, provided the Hilbert space has three or more dimensions.\ft{Bell then  criticizes his own no-hidden-variables argument by challenging the implicit assumption that the result of measuring $\cR$ should not depend on what other jointly measurable physical quantities $\cR$ is measured with. But that's another story.}   

To explain the point of his own refinement of \vN, Bell must explain the problem with \vN's then widely accepted result.  He does this rather informally, condensing  \vN's four assumptions into ``Any real linear combination of any two Hermitian operators represents an observable, and the same linear combination of expectation values is the expectation value of the combination.''  

This overly brisk summary\ft{Bell does mention by name assumptions $B', I,$ and $II$ in a footnote, but says nothing about their separate content.}    insufficiently emphasizes \vN's distinction between physical quantities and Hermitian operators.\ft{Bell does make the distinction in an earlier introductory section, but unlike \vN\ he does not repeatedly insist on it.  His use of ``observable'' to mean ``physical quantity'' is unfortunate, since by 1966 most physicists used the term  for both physical quantities and Hermitian operators.}   
It underemphasizes the importance of the mapping being 1-to-1.    It does not distinguish between assumptions that refer to the statistical Exp functions and assumptions that do not.    Nevertheless, this  rough summary is  enough to make clear what Bell objects to in \vN's assumptions, and this is all he needs to set the stage for his own improvement on \vN.

What Bell objects to is that although the linearity of expectation values of noncommuting operators\ft{Bell  often fails to distinguish between ``noncommuting operators" and ``not jointly measurable physical quantities''.  We show in Appendix 1 that the association of noncommuting operators with physical quantities that are not jointly measurable does indeed follow from assumptions A\', I, and II.  } ``is true for quantum mechanical states, it is required by \vN\ of the hypothetical dispersion free states also.''  But the ``additivity of expectation values $\ldots$ is a quite peculiar property of quantum mechanical states, not to be expected {\it a priori\/}.  There is no reason to demand it individually of the hypothetical dispersion free states, whose function it is to reproduce the {\it measurable\/} peculiarities of quantum mechanics when {\it averaged over\/}.'' [Bell's italics.]

This is the same as the reason we give in Section 2  for the failure of \vN's no-hidden-variables proof: the culprit is indeed assumption B\'.  We have no doubt 
that Bell knew exactly what the problem was.\ft{Bell mentions the nonadditivity of eigenvalues of non-commuting operators not because he thought \vN\ had overlooked this, but because it helps him explain  the ``nontriviality of the additivity of expectation values'', which \vN, unaccountably, takes for granted in the dispersion-free \se s.  Similarly, when Bell mentions that ``a measurement of a sum of noncommuting observables cannot be made by combining trivially the results of separate observations on the two terms'' because ``it requires a quite distinct experiment,''  he is not suggesting that \vN, who returns to this point repeatedly, was unaware of it.}

\medskip
\noindent {\bf 4. Hermann's criticism of \vN.}
\medskip

In 1935, three years after the publication of \vN's book and three decades before John Bell's criticism of that book, Grete Hermann wrote about it.\ft{The part of Hermann, 1935, that we address here is the short (small print!) Section 7, ``The circle in \vN's proof'', pps.~251-253.  Hermann calls \vN\ ``Neumann''.  We have restored his ``von'' for the sake of uniformity.}       She raised the same objection as Bell would thirty years later.      Her criticism of \vN\ is more thorough than Bell's, because she follows \vN's argument more closely.\ft{Perhaps in 1935 the distinctions \vN\ relied on had not yet been absorbed into a terminology that obscured important distinctions.}  By not conflating \vN's four assumptions, she is able to address questions Bell couldn't  formulate (and didn't need to, for his purposes.)  But after precisely identifying \vN's oversight, she offers him some escape hatches that we cannot make much sense of.\ft{We conjecture that she may have found \vN's blatant oversight so surprising that she tried, unsuccessfully, to guess what else he may have had in mind.  It is these final remarks of hers that lead Dieks to state that her views are closer than Bell's to \vN.}

Hermann considers an ensemble of physical systems.  There are physical quantities $\cR$ and $\cS$ 
that can be measured on the systems of the ensemble.   There is a function $\Exp(\cR)$ that gives 
 the mean value of the measurement outcomes arising from an $\cR$-measurement on all the systems of the ensemble.  ``\VN\ assumes that $$\Exp(\cR+\cS) = \Exp(\cR) + \Exp(\cS).\equ(GH-B')$$   In words: {\it the expectation value of a sum of physical
quantities is equal to the sum of the expectation values of the two quantities {\rm [her italics]}:\/}  {\it \vN's proof stands or falls with this assumption.\/} [our italics]"

This crucial assumption is equivalent to \vN's B\'.    It is trivial, Hermann notes, for classical physics, and for  quantum mechanical quantities that can be simultaneously measured, because then 
``the value of their sum is nothing other than the sum of the values that each of them separately takes, from which follows immediately the same relation for the mean values of these magnitudes. The relation is, however, not self-evident for quantum mechanical quantities between which uncertainty
relations hold, and in fact for the reason that the sum of two such quantities is not immediately
defined at all: since a sharp measurement of one of them excludes that of the other, so that the
two quantities cannot simultaneously assume sharp values, the usual definition of the sum
of two quantities is not applicable.  Only by the detour over certain mathematical operators
assigned to these quantities does the formalism introduce the concept of a sum also for such
quantities.''

Hermann  is saying here that because it is not clear how to define  the sum in \(GH-B') or in Assumption B\' of two quantities that are not jointly measurable,  ``to introduce the concept of a sum$\ldots$for such quantities''  requires a detour involving mathematical operators assigned to them --- i.e. \vN's Assumptions I and II.    By emphasizing  the need for a  detour into I and II  she underlines that  it is not necessary to take B\' to {\it define\/} the sum of quantities that are not simultaneously measurable.   
Hermann is reading \vN\ just as we do.\ft{Less anachronistically, we are reading \vN\ just as she does.}

For an ensemble characterized by a wave-function $\phi$, Hermann notes, 
 $$\Exp(\cR) = (R\phi,\phi),\equ(GH-Exp)$$ and therefore \(GH-B') is valid by virtue of the quantum mechanical identity  $$((R + S)\phi, \phi) = (R\phi,\phi) + (S\phi,\phi). \equ(GH-B'')$$   Here $R$ and $S$ are, she notes, ``mathematical operators assigned to the quantities $\cR$ and $\cS$.''   Since \(GH-B'') holds whether or not $R$ and $S$ commute, \(GH-B') holds whether or not $\cR$ and $\cS$ are simultaneously measurable.\ft{In Appendix 1 we prove from \vN's Assumptions A\', I, and II that physical quantities that are  jointly measurable do indeed correspond 1-to-1 to Hermitian operators that commute.}   So B\' does hold for ensembles characterized by wave functions. 

But what about subsets of those ensembles ``selected from them on the basis of any new features.''  
For those  \se s ``one can no longer infer from the asserted addition rule for $(R\phi,\phi)$, that also in these subsets the expectation value of the sum of physical quantities is the same as the sum of their expectation values.    In this way, however, an essential step in Neumann's proof is missing.''   There it is:   precisely the same problem that we describe in Section 2 and that Bell identified thirty years after Hermann.  

\bigskip

  We wish Hermann had stopped here.   But she goes  on.   It is our guess that she goes on because she knows that this obvious problem did not stop \vN.   What can he have been thinking?   At this point we cannot paraphrase her account, because we can no longer  follow it.   We attach it as  Appendix 2, in the hope that the reader may understand her better than we have done.   

Setting aside what we take to be Hermann's  efforts to find the motivation behind  \vN's oversight, she has, in fact, read \vN\ more closely than Bell.   She has the whole story.  Once again, the culprit is Assumption B\'.    The only real difference between the reading we and Bell give and hers, is that she considers the possibility that \vN\  himself was aware of the obvious problem, and implicitly limited himself to \se s for which the difficulty did not arise.    But if he did that, then he had committed himself to the view that the hidden variables single out only those subensembles that  lack features which make them any different from the larger $\phi$-ensembles that they combine to give.  So even if he did know what he was doing, he was begging the question. 

\bigskip
\noindent {\bf Acknowledgment}
\medskip
We would like to thank Ulrich Mohrhoff for bringing Dieks, 2017, to our attention.

\bigskip
\noindent {\bf Appendix 1}
\medskip

Here are two examples of the power of \vN's Assumptions A\', I, and II.   The problematic assumption B\' is not used;   sums of physical quantities that are not jointly measurable are defined by sums of the corresponding Hermitian operators, using Assumption II.  See also \vN, 1932, and Park and Margenau, 1968.  
\bigskip
\noindent {\bf Theorem:}  {\sl The result of measuring a physical quantity must lie in the spectrum of the corresponding Hermitian operator.} 
\medskip 

Let $R$ be the Hermitian operator corresponding to the physical quantity $\cR$, and consider a function $f(x)$ that is 0 if $x$ belongs to the spectrum of $R$, and 1 elsewhere. This means that $f(R)$ = 0, the zero operator. So Assumption I requires that  $f(\cR) = 0,$ the physical quantity that is always 0.
Thus, for every result $r$ of measuring $\cR$, we have $f(r)=0$, which means that $r$ belongs to the spectrum of $R$.

\bigskip
\noindent {\bf Theorem:}  {\sl  The correspondence between physical quantities and Hermitian operators must associate jointly measurable quantities with commuting operators and vice-versa.}
\medskip

If two Hermitian operators $R$ and $S$ commute, then it is a mathematical fact that there is a third Hermitian operator $T$ of which they are both functions: $$ R = f(T),\ \ S = g(T).\equ(A2.1)$$
By Assumption I the correspondence preserves functional relations, so 
$$ \cR = f(\cT),\ \ \cS = g(\cT).\equ(A2.2)$$   Using Assumption A\', one can then simultaneously measure $\cR$ and $\cS$ by measuring $\cT$ and applying to the outcome of that measurement the functions $f$ and $g$.  

\medskip
The converse is trickier.\ft{
Unlike the proofs in  \vN, 1932, or Park and Margenau, 1968,  the elementary, but somewhat complicated, proof that follows  
makes no use of the spectral decomposition theorem for Hermitian operators.}    Let the physical quantities $\cR$ and $\cS$ be jointly measurable, and let $R$ and $S$ be the corresponding Hermitian operators.  By Assumption A\', functions of such jointly measurable quantities can be measured by measuring the individual quantities $\cR$ and   $\cS$ and evaluating the function at the individual outcomes $ r$ and  $s$.     So products of two jointly measurable physical quantities, that differ only in the order in which the quantities appear, can all be measured by the same experiment and have the same measurement outcomes.   Such physical quantities are therefore identical.  Since the correspondence is 1-to-1, they must therefore all correspond to the same Hermitian operator.   We show below that this leads to identities among the operators $R$ and $S$ which we can exploit to show that $R$ and $S$ must commute.   

To begin with, it follows from Assumptions I and II  that the physical quantity $(\cR + \cS)^2 - \cR^2 - \cS^2 = 2\cR\cS = 2\cS\cR$ corresponds to the Hermitian operator $(R+S)^2  - R^2 - S^2 = (RS+SR)$.  Therefore
the Hermitian operator corresponding to both $\cR\cS$ and $\cS\cR$ when $\cR$ and $\cS$ are jointly measurable is given by
$$\cR\cS\  = \ \cS\cR\ \ \ \llra \ \ (RS+SR)/2. \equ(symprod)$$

The next step is to apply the general rule \(symprod) to another jointly measurable pair, $\cR$ and  $\cR\cS$: $$\cR(\cR\cS) \ \ \llra \ \  [R(RS+SR)/2 + (RS+SR)R/2]/2 = (R^2S + 2RSR + SR^2)/4.\equ(RRS)$$
One more application of \(symprod), to the pair $\cS$ and $\cR(\cR\cS)$, gives 
$$\cS(\cR(\cR\cS)) \ \ \llra \ \ [2(SR)(RS) + 2(SR)^2 + 2(RS)^2 + S^2R^2 + R^2S^2 ]/8.\equ(SRRS)$$
Interchanging the names of $\cS$ and $\cR$ we also have
$$\cR(\cS(\cS\cR)) \ \ \llra \ \ [2(RS)(SR) + 2(RS)^2 +  2(SR)^2 + R^2S^2 + S^2R^2]/8.\equ(RSSR)$$

On the other hand, directly squaring both sides of  \(symprod) gives
$$(\cR\cS)^2 \ \ \llra \ \  [(RS)^2 + (SR)^2  + (RS)(SR) + (SR)(RS) ]/4. \equ(RSRS)$$
Since $\cS(\cR(\cR\cS))$, $\cR(\cS(\cS\cR))$, and $(\cR\cS)^2$ are all the same physical quantity, the sum of the right sides of \(SRRS) and \(RSSR) must be the same operator as twice the  right side of \(RSRS).  This gives us 
$$R^2S^2 + S^2R^2 = (RS)(SR) +  (SR)(RS) .\equ(R2S2)$$  
We also have, as a direct application of  \(symprod) to the pair $\cR^2$ and $\cS^2$,   $$\cR^2\cS^2 \ \ \llra\ \ [R^2S^2 + S^2R^2]/2,\equ(SSRR)$$
and therefore, in view of \(R2S2), 
$$\cR^2\cS^2 \ \ \llra \ \  [(RS)(SR) + (SR)(RS)] /2.\equ(SSRR)$$
Since the operators on the right sides of \(SSRR) and \(RSRS) must be the same, we have
$$(RS)(SR) + (SR)(RS)  = (RS)^2+(SR)^2, \equ(comm2)$$ and therefore
$$(RS-SR)^2 = 0.\equ(comm3)$$

This requires the square of the Hermitian operator $C = i(RS-SR)$ to vanish, which in turn requires $C$ itself to vanish.  So $R$ and $S$ must indeed commute.

\medskip
\noindent {\bf Appendix 2}
\medskip

We reproduce below the final paragraph and a half of the section on \vN\ in Hermann, 1935.  The footnotes are our own comments.  We have a sense of what Hermann is trying to say in the first half-paragraph below and the first half of the last paragraph.   We have no useful comments  on the final part of the final paragraph.  We would guess that she is struggling, unsuccessfully, to guess the kind of thinking that led to \vN's surprising oversight.

{{\narrower

In this way, however, an essential step in \vN's  proof is missing.\ft{The step is being able to apply B\' to the 
($\phi,\lambda$)-\se s extracted from a $\phi$-ensemble by the value $\lambda$ of the hidden variables.}       If instead --- like \vN\ --- one does not give up on this step,\ft{Here she notes that \vN\ does take the step.}  then one has implicitly absorbed into the interpretation the unproven assumption that there can be no distinguishing features, of the elements of an ensemble of physical systems characterized by $\phi$, on which the result of the $\cR$-measurement depends.\ft{She is saying that since he does take the step he must be implicitly assuming that the ($\phi,\lambda$)-subensembles have no features to distinguish them from the original $\phi$-ensemble, for which B\' is valid.}   However, the impossibility of such features is precisely the claim to be proven.  Thus the proof runs in a circle.\ft{She allows \vN\ the option of begging the question, rather than overlooking  an obvious objection.}

On the other hand, from the standpoint of \vN's calculus one can argue against this,
that  it is an axiomatic requirement that all physical quantities are uniquely
associated with certain Hermitian operators in a Hilbert space, and that through the discovery
of new features invalidating the present limits of predictability, this association would
inevitably be broken.\ft{Here she echoes \vN\ in suggesting that the failure of  I and II might  be behind the existence of a dispersion free subensemble.}  Indeed, any discovery that is representable in the operator calculus
would have its contents specified only through the form of a wave function, which for quantities
not simultaneously measurable exhibits the smearing out required by the uncertainty
relations, and which finds application only by way of the probability interpretation.   By this consideration, however, the crucial physical question of whether the progress of physical research can attain more precise predictions than are possible today, cannot be twisted into the impossibly equivalent mathematical question of whether such a development would be representable solely in terms of the quantum mechanical operator calculus.  There would need to be a compelling physical reason, if not only the physical data known to date, but also all the results of research still to be expected in the future are related to each other according to the axioms of this formalism. But how should one find such a reason?
The fact that the formalism has so far proven itself, so that one is justified in seeing in it
the appropriate mathematical description of known natural connections, does not mean that
the as yet undiscovered natural law connections should also have the same mathematical
structure.

}}

\bigskip
\noindent {\bf References}

\medskip\ni
Bell, J. S., 1964, ``On the Einstein-Podolski-Rosen paradox,'' Physics {\bf 1}, 195--200.   Reprinted in Bell, 1987.
\medskip\ni
Bell, J. S., 1966, ``On the problem of hidden variables in \qm", Reviews of Modern Physics {\bf 38}, 447--452.  Reprinted in Bell, 1987.
\medskip
\ni
Bell, J. S., 1987, {\it Speakable and Unspeakable in Quantum Mechanics\/},
Cambridge University Press, Cambridge.
\medskip
\ni
Bub J., 2010. ``Von Neumann's `no hidden variables' proof: A re-appraisal,'' Foundations of Physics, {\bf 40},  1333--1340; https://arxiv.org/abs/1006.0499.
\medskip
\ni
Bub J., 2011. ``Is Von Neumann's `no hidden
variables' proof silly?", Chapter 10 of ``Deep beauty --- 
understanding the quantum world through mathematical innovation'', H. Halvorson, ed., Princeton University Press, 
393--407.
\medskip
\ni
Dieks, D., 2017, ``Von Neumann's impossibility proof: Mathematics
in the service of rhetorics'', Studies in the History and Philosophy of Modern Physics {\bf 60},    136--148;
https:// arxiv.org/abs/1801.09305. 

\medskip
\ni
Hermann, G., 1935, ``Die naturphilosophischen Grundlagen der Quantenmechanik
(Aus\-zug), ``Abhandlungen der Fries'schen Schule {\bf 6}, 75-152.  English translation:  Chapter 15 of 
``Grete Hermann --- Between physics and philosophy", Elise Crull and Guido Bacciagaluppi, eds., Springer, 2016, 239--278.    [Volume 42 of Studies in History and Philosophy of Science]

\medskip
\ni
Mermin, N. D., 1993, ``Hidden variables and the two theorems of John Bell,''
Reviews of Modern Physics {\bf 65}, 803--815.  
Recently posted at https://arxiv.org/abs/1802.10119; three minor errata have been repaired, some footnotes of commentary have been added, and the present manuscript is announced as forthcoming.     

\medskip
\ni  Park, J.~L., and H. Margenau, 1968, ``Simultaneous measurability in quantum theory," 
International Journal of Theoretical Physics, {\bf 1}, No. 3, 211--283.

\medskip 
\ni
von Neumann, J., 1932 {\it Mathematische Grundlagen der Quantenmechanik\/},
Springer-Ver\-lag, Berlin.  English translation:\ \  {\it Mathematical Foundations
of Quantum Mechanics\/}, Princeton University Press, Princeton, N.J., 1955.

\bye